\documentclass[12pt,letterpaper]{article}
\usepackage{amsmath,amssymb,pgf,pgfarrows,pgfnodes,float,appendix, hyperref}
\usepackage{graphicx}
\usepackage{subfigure}
\usepackage[margin=0.9in]{geometry}

\newcommand{\be}{\begin{equation}}
\newcommand{\ee}{\end{equation}}
\newcommand{\bea}{\begin{eqnarray}}
\newcommand{\eea}{\end{eqnarray}}

\title{{\rm\footnotesize \qquad \qquad \qquad \qquad \qquad \ \qquad \qquad \qquad \ \ \ \ \ \                  RUNHETC-2016-15, UTTG-12-16 }\vskip.5in    Holographic Space-time Models of Anti-deSitter Space-times }
\author{Tom Banks\\
Department of Physics and NHETC\\
Rutgers University, Piscataway, NJ 08854\\
E-mail: \href{mailto:banks@physics.rutgers.edu}{banks@physics.rutgers.edu}
\\
\\
Willy Fischler\\
Department of Physics and Texas Cosmology Center\\
University of Texas, Austin, TX 78712\\
E-mail: \href{mailto:fischler@physics.utexas.edu}{fischler@physics.utexas.edu}}
\date{}
\begin{document}
\maketitle

\begin{abstract}
We study the constraints on HST models of AdS space-times, beginning from a general formalism for FRW models.  The causal diamonds of HST along time-like geodesics of AdS space-time, fit nicely into the FRW patch of AdS space.  The coordinate singularity of the FRW patch is identified with the proper time at which the Hilbert space of the causal diamond becomes infinite dimensional.  For diamonds much smaller than the AdS radius, $R_{AdS}$, the time dependent Hamiltonians of HST are the same as those used to describe similar diamonds in Minkowski space.  In particular, they are invariant under the fuzzy analog of volume preserving mappings of the holographic screen - unitary rotations of the fundamental spinor components.  This leads to fast scrambling of perturbations on the horizon of a black hole of size smaller than  $R_{AdS}$.  We argue that, in order to take a limit of this system which converges to a CFT, one must choose Hamiltonians, in a range of proper times of order  $R_{AdS}$, which break this invariance, and become local in a particular choice of basis for the variables.
We show that, beginning with flat, sub-$R_{AdS}$, patches of dimension $D$, the resulting CFT, constructed from the variables of HST,  is inconsistent with the entropy of large black holes, {\it unless} one has at least two compact dimensions, whose size is of order  $R_{AdS}$. The argument that at least two of the compact dimensions are large is connected to a new observation about the scrambling rate of information localized on the compact dimensions. Our construction explains, in a natural manner, why large AdS black holes do not have the fast scrambling property.  Our present approach cannot deal with models where string theory is weakly coupled and  $R_{AdS}$ is of order the string scale, because the relationship between area and entropy is non-universal in such models. On spatial length scales longer than  $R_{AdS}$, our mapping of  HST variables into CFT shares much with the Tensor Network Renormalization Group (TNRG){\cite{tnrg}} and is a sort of evolving error correcting code\cite{code}.
\end{abstract}

\section{Introduction}

Some time ago we showed how models based on the principles of Holographic Space Time (HST) could reproduce a number of features expected from a quantum theory of gravity\cite{previous}.  The other non-perturbative approaches to quantum gravity, which we consider valid, are Matrix Theory\cite{bfss} and the AdS/CFT correspondence\cite{adscft}, and it is interesting to ask what the relationship is between HST and these more traditional approaches.  In this paper we will accomplish that goal, at least at the level of general scaling properties, for AdS/CFT .  The conventional approach to AdS/CFT emphasizes three regimes for the ratio $x \equiv R_{AdS}/L_P$.  For $x$ of order $1$ there is no space-time interpretation of the CFT at all.  For $x \gg 1$, there is {\it usually} a regime well described by a bulk theory of gravity.   In weakly coupled string models, there is an intermediate range of $x$, determined by the ratio $l_S / L_P \gg 1$.  Bulk gravity is only a good approximation when $x \gg l_S / L_P $.  One way to understand this is to note that the effective gravitational action has higher dimension operators scaled by powers of $l_S$ rather than $L_P$, the D-dimensional Planck scale $(=1)$.  Jacobson\cite{ted2} has shown that such an action can be derived as the hydrodynamic equation of a system in which the relation between the area of a holographic screen and the entropy is modified from the Bekenstein-Hawking-'t Hooft-Fischler-Susskind-Bousso  
law, by terms involving powers of $\frac{l_S^{d-2}}{A}$.  Since the coefficients of the effective string theory action are model dependent\footnote{{\it e.g.} in compactifications involving nontrivial 1-cycles, around which strings can wind, the coefficients depend on the sizes of the cycles in $l_S$ units.} , the HST formalism is stymied because there is no universal relation between area and entropy.  For this reason, in this paper we will restrict attention to models where $g_S \sim 1$, or to M-theory or F-theory models, where the only scales are geometrical scales $\gg 1$.  We note that a similar simplification of the comparison of HST with existing string theory models at strong coupling was found in the $1 + 1$ dimensional models of \cite{tb 1 + 1}.  

In the next section, we begin our discussion by showing how the description of HST in the FRW patch of AdS space, leads to structures resembling TNRG.  The renormalization group time step in TNRG can be viewed as implemented by a time dependent Hamiltonian coupling together more and more of the variables of an underlying lattice model\cite{tnrg}.  This is precisely what happens in HST as a function of proper time, except that the direction of RG flow is opposite to the direction of increase of the proper time interval.  Moreover, the variables of HST in a causal diamond with small proper time are the sections of the spinor bundle on the $ d - 2$ dimensional holographic sphere, with an eigenvalue cutoff on the Dirac equation (equivalent on the sphere to an angular momentum cutoff) , so we are indeed approaching a fixed point as the time is increased.   The main difference between the two approaches is that HST works with an angular momentum cutoff, while TNRG works with a lattice cutoff.  We make a connection between the two by introducing a {\it fuzzy lattice}, defined in terms of the cutoff delta function on the fuzzy sphere.  In this way, we will see that the momentum space cutoff is a kind of error correcting code\cite{code}.  The other major difference between TNRG/error correction and HST is that the former is usually visualized on a fixed HST time slice.  The fictitious time parameter of TNRG has not previously been identified with an actual time in AdS space, probably because in that approach one is always thinking about a global time slice, rather than the causal diamond slices of HST.  A similar distinction between HST and global FRW time slices was crucial to our HST model of inflation\cite{holoinflation2} .  

The above procedure, generically produces an AdS space with a radius of order Planck scale.   By this we mean that there is no regime of proper time in which we can see meta-stable excitations with properties approximately the same as those of flat space black holes.  In order to force the existence of such a regime, we insist that there is a period of proper time $\ll R_{AdS} $ during which the time dependent Hamiltonian is that of a $D$ dimensional Minkowski space.  We've shown how to construct such Hamiltonians, which do have excitations resembling black holes, and others resembling jets of massless particles\cite{holonewton2}.
In particular, the ``small" black holes of these models scramble perturbations of their horizons on a time scale of order $R {\rm ln}\ R$, where $R$ is the black hole radius.  This is {\it not} the case for large AdS black holes, which are thermal states in field theory on the holoscreen, and scramble information ballistically\footnote{We thank S. Shenker for a discussion, which corrected our statement that all information was scrambled diffusively in quantum field theory.} ({\it i.e.} in a time of order $R$) and information about conserved quantum numbers diffusively.

To model a large radius AdS space then, we must precede the period of TNRG evolution described above, by a period of Minkowski evolution. If we simply do this, in $D$ dimensions, we reach a contradiction with the properties of large black holes in AdS space.  The CFT we produce by this construction has a finite temperature free energy
$ F = c (T R_{AdS})^{D - 1} $, with $c$ of order $1$.  To avoid this contradiction, we must instead consider only $d \leq D - 1$, of the original dimensions to form an $AdS_d$, while $D - d$ dimensions are compact, with volume $R_{AdS}^{D - d}$.  The necessity for large radius AdS spaces to be accompanied by large compact dimensions is a familiar ``phenomenological" fact about known AdS/CFT correspondences.  HST provides an explanation for this fact.  In fact, the ``phenomenology of the AdS landscape of String Theory", suggests that there are consistent models only with $D - d \geq 2$.  We provide an explanation of this fact as well, by arguing from classical GR that large black holes in $AdS_d \times {\cal K}_{D - d}$ fast scramble information localized on the compact dimensions.  Our HST models easily accommodate this fast scrambling, but only for $D - d \geq 2$.

\section{HST in FRW Space-time}

Open FRW space-times with negative curvature have a metric of the form
\begin{equation} ds^2 = - dt^2 + a^2 (t) (\frac{dr^2}{1 + \frac{r^2}{R_{AdS}^2}} + r^2 d\Omega_{d - 2}^2) , \end{equation} where 
$d\Omega_{d - 2}^2 $ is the metric on the round $ d - 2$ sphere.  This is the same as
\begin{equation} ds^2 = a^2 (\eta) (- d\eta^2 + \frac{dr^2}{1 + \frac{r^2}{R_{AdS}^2}} + r^2 d\Omega_{d - 2}^2 ) , \end{equation} where the conformal time $\eta$ is the solution of
\begin{equation} d\eta = \frac{dt}{a(t)} . \end{equation}  We will take $a(t) = \cos (\frac{\pi t}{2 R_{AdS}})$, which is time symmetric.  We take $\eta (t = 0 ) = 0$. The causal diamond at time $\eta$, along the geodesic at $r = 0$ has its holoscreen at $ t = 0$, with $r = R_{AdS} arcsinh (\eta / R_{AdS})$. The conformal time $\eta$ approaches infinity as $t$ approaches $R_{AdS}$, so at this time the area of the holoscreen is infinite (Note that $a(0) = 1$, so the area is just $r^{d-2}$ times the area of a unit sphere.).  The Hamiltonian evolution after $t = R_{AdS}$ takes place in this infinite dimensional Hilbert space, and is given by the time independent Hamiltonian of some CFT.  Our goal in this paper is to understand how the local evolution matches onto the CFT evolution as $t$ approaches the singularity of $\eta$.  Note that in order to discuss causal diamonds of size less than $R_{AdS}$, we must restrict the proper time by the inequality $arcsinh (\eta (t)/ R_{AdS}) <1$. The time dependent Hamiltonian of HST should be close to that for Minkowski space \cite{holonewton2} for times satisfying this inequality, and then evolve to $H_{CFT}$ as the singularity is approached.  

We need to say a word about our choice of coordinates.  We choose the FRW coordinates on AdS because they cover the maximal causal diamond of a single time-like trajectory, over a proper time interval for which the diamond's area is finite.  We interpret the coordinate singularity at $t = R_{AdS}$ as the appearance of an infinite dimensional Hilbert space at this time.  As we approach the singularity, the Hamiltonian must make a transition between the fast scrambling behavior characteristic of flat space black holes, and the ballistic/diffusive behavior of quantum field theory.  We will argue that starting at a time of order the AdS radius, the fast scrambling Hamiltonian is replaced by a sequence of Hamiltonians converging on $H_{CFT}$, in a manner resembling the (backward) renormalization group flow of TNRG.  The local, approximately Poincare invariant, physics of the diamond is NOT captured by the TNRG.

The actual time slices followed by our quantum theory of AdS space, are not the FRW slices, but ``causal" slices fitting between causal diamonds whose tips are separated by a Planck time.  Each slice coincides with the corresponding FRW slice at the position on which the trajectory pierces it.

A cautionary note:  Our FRW coordinate system could be placed so that $t = 0$ coincides with {\it any} value of the boundary global time.   It is a useful description of a physical process in the CFT, if an operation on the CFT ground state can produce a state on the past boundary of the FRW causal diamond, which corresponds to some local excitation of a corresponding region in empty Minkowski space.  We will later discuss what is involved in constructing such an operator.
The discussions of \cite{lennyjoeetal} are essentially the claim that one {\it can} construct such an operation for scattering amplitudes involving a small number of particles with energies well below the Planck scale.  

In the FRW coordinate system, along a given geodesic, the equation for area as a function of proper time instructs the practitioner of HST to associate a Hilbert space ${\cal H}_{in} (t)$ with the corresponding causal diamond.  The general rules of HST tell us that this Hilbert space is the minimal representation of some super-algebra, whose fermionic generators are elements of the spinor bundle over the holoscreen.  For area large in Planck units, the Hilbert space has large entropy, $S$ and this entropy comes from keeping fermionic generators with angular momentum up to $L \sim S^{\frac{1}{d-2}}$.  For simplicity, we will assume that the anti-commutation relations of the generators are
\begin{equation} [\psi_l , \psi^{\dagger}_{k} ]_+ = \delta_{kl} . \end{equation} Here $l$ and $k$ are multi-indices, indicating both the value of the total angular momentum up to $L$, and the spinor index, as well as the eigenvalues of different Cartan generators of $SO(d-1)$.  We can define an approximate local field by
\begin{equation} \psi_L (\Omega ) = \sum_l^L \psi_l Y_l (\Omega ) , \end{equation} where $Y_l$ are the corresponding spinor spherical harmonics.   The anti-commutator of the field at two different points is 
\begin{equation} [\psi_L (\Omega) , \psi_L^{\dagger} (\Theta ) ]_+ = \sum_l Y_l (\Omega ) Y^*_l (\Theta ) \equiv \delta_L (\Omega, \Theta ) , \end{equation} which we call the fuzzy delta function.  We claim that we can find a collection of points $\Omega_i$ such that 
\begin{equation} \delta_L (\Omega_i, \Omega_j ) = 0 . \end{equation}   This is obvious for $d = 3$, but true for other values of $d$ as well.  The collection of points can be made invariant under discrete subgroups of the Cartan torus of $SO(d - 1)$ but not under non-abelian subgroups.  There are thus many such collections.  For each value of $L$, corresponding to a fixed $r (\eta (t))$ for the holographic screen, we can choose one point on each Cartan orbit in the collection to coincide with a point on the Cartan orbit of the next smaller sphere.  It's not clear whether this is the most useful convention, but we will choose it.  This defines a set of finer and finer {\it fuzzy lattices} on the nested spheres.  It's important to note that the fields $\psi_L $ and $\psi_K$ on the same point are {\it not} the same operator, because for $K > L$ $\psi_K$ will contain more angular momentum modes.  We can express $\psi_L$ in terms of many different linear combinations of $\psi_K ( \Omega_i )$, where $\Omega_i$ are all the points in the lattice of fineness $K$ .  Essentially we are embedding a vector space as a hyperplane in a vector space of higher dimension, in terms of a basis whose vectors are not parallel to the hyperplane.  Thus, given a state of the system of $\psi_K (\Omega_i )$, we can change the state in many ways, which do not change information stored in the $\psi_L$.  Thus, the description of our system in terms of fuzzy lattices has the robustness of IR information, which was characteristic of the error correcting codes of \cite{code} .  The code in our case is time dependent, while the picture in \cite{code} is at a fixed global time.  

\subsection{HST and TNRG}

The Hamiltonian $H_{in} (t)$ acts on the Hilbert space generated by the $\psi_L (\Omega)$ with $L \propto r (\eta (t))$.  We choose discrete time steps (which are of order Planck scale) so that the highest angular momentum operator was previously decoupled from the system.  The evolution is unitary because in the large Hilbert space that we converge to as $t \rightarrow R_{AdS}$ we have
\begin{equation} H (t) = H_{in} (t) + H_{out} (t) .  \end{equation}  Thus the process of time evolution is one in which more and more of the variables of the full system become entangled.  The reverse evolution, shrinking the proper time to $s R_{AdS}$ with $s < 1$, resembles a TNRG\cite{tnrg} transformation, which coarse grains the degrees of freedom.  The TNRG constructs a sequence of Hamiltonians acting on smaller and smaller Hilbert spaces, and when TNRG is applied to a lattice system at a critical point, it has been shown that these small matrices have eigenvalues which agree well with the dimensions of low dimension operators in the CFT. 

If we had a TNRG construction of the CFTs relevant for AdS/CFT we could use the procedure to describe the proper time evolution in HST for times of order 
$s R_{AdS}$ with $s \lesssim 1$.  When $s = 1$ we would hit the fixed point theory, and the evidence is that the TNRG procedure would have converged to the CFT Hamiltonian.  From that point on the Hamiltonian is time independent and we are doing continuum quantum field theory.  

There are a number of technical problems that would have to be solved in order to implement this procedure.  Usually TNRG is formulated in terms of Hamiltonians on a regular lattice.  We see no particular problem of replacing a regular lattice by a fuzzy lattice on a sphere.  Note however that this procedure breaks the $SO(D - 1)$ rotational subgroup that preserves the time-like geodesic.  This invariance is then restored at the fixed point. In all of our previous work, we have taken great care to preserve this symmetry and breaking it would seem to be at odds with what a geodesic observer would see in AdS space.  We can restore the symmetry by integrating the Hamiltonian over all rotational images of the fuzzy lattice, but the resulting Hamiltonian would have power law non-locality on the fuzzy lattice, because it has a sharp angular momentum cutoff.  It's not clear that this Hamiltonian would actually converge to $H_{CFT}$.  A more fundamental problem is that we do not have anything like a TNRG formulation of any of the strongly coupled CFTs relevant to large radius AdS space.  

\subsection{Overlap Conditions and $SO(1,D-1)$ Invariance}

So far we've described the system from the viewpoint of a single geodesic.  Now we consider another geodesic at rest in the FRW coordinates, and related to the first by a spatial isometry of the FRW slice.  If the generators of $SO(2,D-1)$ are 
$L_{\mu\nu}$, with the $D$ and $(D + 1)$st index referring to the two time-like directions, then the generators of hyperbolic rotations on the FRW slice are either $L_{Di}$ or $L_{D + 1\ i}$, where the choice depends on the choice of the arrow of global time.  

When we consider the overlaps of two equal area diamonds, centered at $t = 0$
then the size of the overlap diamond is zero for small enough proper time and approaches infinity as the proper time approaches $R_{AdS}$.  The overlap's fraction of the full diamond area increases with time and becomes greater than $1/2$.  Page's theorem tells us that once this occurs, the states in the individual trajectories' diamonds are almost completely determined by the density matrix on the overlap.  In the limit, the overlap fraction goes to $1$ and the entanglement is complete. The fundamental consistency condition of $HST$, that the individual states give rise to density matrices for the overlap with equal eigenvalues, then implies that the quantum states in the two diamonds are related by a unitary transformation.  $H_{CFT}$ does not commute with this $SO(1,D- 1)$ group, since no two geodesics are at rest in the same static coordinate system.  Thus, these two emergent symmetries close on the full conformal group.

\section{Small Black Holes, Large Black Holes, and Large Compact Dimensions}
\subsection{Small Black Holes}

The above construction, which resembles a TNRG calculation of the properties of a CFT, gives no hint of local structure on scales smaller than the AdS radius.
In particular, there is no hint of the existence of meta-stable excitations with the properties of {\it small black holes}, as we would expect in a model with $R_{AdS} \gg 1 $.  We conclude from this that if we combine the area law with the RG construction of time dependent Hamiltonians approximating $H_{CFT}$, we will obtain a model with $R_{AdS} \sim 1$.  In order to have a large AdS radius, we must insist that for a proper time $\sim R_{AdS} \gg 1 $ we have a Hamiltonian which is approximately that of a Minkowski space model.  Let us assume that we have $D$ Minkowski dimensions.  In \cite{holonewton2} we presented a class of models with just this property.  To be frank, most of those models do not satisfy the constraints of Lorentz boost invariance, but that will not affect the considerations of this paper.  In particular, all of the models of \cite{holonewton2} contained meta-stable excitations with the parametric relations expected for black holes in $D$ dimensions.  The restoration of equilibrium in these black hole ensembles is very rapid, consistent with the expectations of {\it fast scrambling}\cite{hpss}.  Intuitively, the reason for this is the following:  the Hamiltonian for the internal dynamics of a black hole, on time scales less than its evaporation time is
\begin{equation} H_{BH} = \frac{1}{R} {\rm tr}\ P(\frac{M}{R^{d - 3}} ) . \end{equation}  $R$ is the black hole radius in units, and \begin{equation} M_i^j = \psi_{i,a_1 \ldots a_{D  - 3}} \psi^{\dagger\ a_1 \ldots a_{D - 3}, j} . \end{equation}  The $\psi$ variables are elements of the spinor bundle on the holographic screen of the black hole, which is a $D - 2$ sphere on its horizon.  The spinor bundle has an angular momentum cutoff of order $R$ and all indices are anti-symmetrized and run from $1$ to $R$. This Hamiltonian is invariant under a large group of unitary transformations, which approximates the group of volume preserving mappings on the sphere.  Thus, if we choose a basis of localized functions on the sphere, information encoded in the state of one localized variable, is actually in interaction with all of the rest of the sphere, because a small spherical cap is equivalent to a $D - 2$ dimensional amoeba of equal volume.  We have conjectured, following the work of Sekino and Susskind\cite{hpss} for the case $D = 4$ that a generic model of this type will have fast scrambling behavior.  

If we look at the black hole in a causal diamond with proper time $T$ in the range $R^{D - 1} \gg T \gg R$, then the full Hamiltonian has the form
\begin{equation} H_{in} (T) = P_0 + \frac{1}{T} {\rm tr}\ P(\frac{M}{T^{d - 3}} ) . \end{equation} The variables appearing in the second term now have indices running from $1$ to $T$, but the initial state of the system on the past boundary of the diamond is constrained by the equations
\begin{equation} \psi_{A, b_1 \ldots b_{D - 3}} | BH \rangle = 0 . \end{equation} The $b_i$ run from $1$ to $R$ and $A$ runs from $R + 1$ to $T$ .   This constraint decouples the variables associated with the black hole, from the rest, which are thought of as ``almost gauge modes" on the holographic screen of the large diamond\cite{holonewton2}\cite{bms}, for proper times of order $T$.   $P_0 =  \frac{1}{R} {\rm tr}\ P(\frac{M_R}{R^{d - 3}} )$, where $M_R$ is the matrix constructed from the $\psi_{b_1 \ldots b_{D - 2}}$.  The energy of a typical state of this Hamiltonian is $\sim R^{D - 3}$, while the natural scale for energy splittings is $o(1/R)$.
Of course, since the system has entropy $S \sim R^{D - 2}$ there are also splittings as small as $e^{-S} / R$, but these are irrelevant on scrambling or evaporation time scales.

 If there are additional particles or black holes inside the diamond, there will be more constraints on the initial state, which guarantee that those objects behave like isolated localized systems, for time scales less than $T$.  {\it The fact that bulk localized objects in Minkowski space, correspond to constrained states of variables located on the holographic screen, is the most important lesson we have learned from HST}.  

If the Minkowski like diamonds described above are actually living in an FRW patch of AdS, then when $T \sim R_{AdS}$ we must abandon the above prescription for $H_{in} (T)$ and replace it with the TNRG prescription of the previous section.  Our first task is to understand how the small black hole excitations described above, appear in the CFT.  The Hilbert space in which they had their transient existence is a tensor factor in the CFT Hilbert space, so we must be able to find states of the same energy, and entropy in the conventional description of the CFT Hilbert space.   Qualitatively, intuitions about Hawking radiation tell us what happens: small black holes eventually settle down to states of the AdS gas.   Are there enough such states, with the right properties?
Our picture of the small black hole states in terms of HST in Minkowski space, has the following properties as a function of the black hole radius: the typical energy is $R^{D - 3}$ and the entropy, $R^{D - 2}$.  The scale of typical splittings is $1/R$.\footnote{This is the generalization of the 't Hooft scaling law for splittings in large $N$ matrix models to the tensor models we need in $D > 4$ dimensions.}  In CFT language, we generate states by acting with operators on the conformally invariant state.  The energy {\it in units of $R_{AdS}^{-1}$ } of a product of operators acting at different points on the sphere is the sum of their dimensions.   If we choose to act with a few low dimension operators at the $R^{D - 2}$ points on the fuzzy lattice appropriate to a black hole of radius $R$ we get an entropy of order
$R^{D - 2}$ and an energy of order $\frac{R^{D - 2}}{R_{AdS}}$.  The energy is thus too small by a factor of $\frac{R}{R_{AdS}}$.  The same is true of the splittings, which are typically of order $\frac{1}{R_{AdS}}$ in the CFT.  It does not seem possible to explain these discrepancies in terms of a red shift between the static and FRW coordinates, since the redshift depends quadratically on  $\frac{R}{R_{AdS}}$.  

Instead we believe this indicates that the Hamiltonian describing the finite time local evolution of small black holes, is {\it not} $H_{CFT}$, but rather the time dependent $H_{in} (T)$ for $ T < R_{AdS}$.  There are many other reasons to draw this conclusion.  Chief among these is the fact that field theory Hamiltonians, either CFT or cut-off, do not fast scramble information localized in angle on the $ D - 2$ sphere.  In addition, the states of small black holes in randomly chosen FRW diamonds, are extremely complicated when expressed in terms of low energy states of the CFT.  Our CFT is realized as a UV limit of a Lagrangian for cut-off fermion fields.  We also use the fact, well known from the AdS/CFT correspondence, that a CFT with a large radius dual is not close to a free field theory.  It follows that the eigenstates of the CFT will, by Page's theorem\cite{page}, have maximal entanglement between the tensor factor of the Hilbert space on which only fermions with angular momentum $< R_{AdS}$ act, and the rest.  Thus, in CFT eigenstates, with probability very close to one, the density matrix in the tensor 
factor corresponding to an FRW diamond is proportional to the unit matrix.   The probability of having a black hole of radius $R$ in an FRW diamond is thus $e^{- R^{D - 3} R_{AdS}}$.   Another way to see this is that the equilibrium state into which a small black hole decays is the AdS gas, which has very tiny probability to spontaneously recreate a localized black hole.  

When the CFT Hamiltonian acts on the small black hole state, it has no reason to create states in which the tensor factor of fermions with angular momentum $< R_{AdS}$ remains unentangled with higher angular momentum modes, which represent the region of AdS space outside the FRW diamond.  Thus, it cannot mimic the action of $H_{in} (t)$ for $- sR_{AdS} < t < sR_{AdS}$, $s = \frac{e - e^{-1}}{2}$.  The fact that we {\it are} able to match the energy and entropy of the small black holes of radius $\sim R_S$ to identifiable states in the CFT, suggests that it is possible to make a smooth extrapolation of $H_{in} (t)$ so that it connects to the TNRG Hamiltonians and converges to $H_{CFT}$ as $t$ approaches the zero of $a(t)$.  Note however, that in order for the scrambling process to be undisturbed by the presence of AdS curvature, we must have $R {\rm ln}\ R < R_{AdS}$, so that we are not quite in the regime where the match can be made.  If we want to study the full flat space evaporation process in AdS/CFT, then we must have $R^{D - 1} < R_{AdS}$, and we certainly do not expect the CFT time evolution to give a correct description of the problem.

\subsection{Large Black Holes}

We now turn to the large stable black holes in $AdS_d$ space.  The metric of these states in static coordinates is
\begin{equation} ds^2 = - dt^2 f(r) + \frac{dr^2}{f(r)} + r^2 d\Omega_{d-2}^2 . \end{equation}  In this equation
\begin{equation} f(r) = (1 - (\frac{ R_0}{r})^{d - 3} + \frac{r^2}{R_{AdS}^2} ) . \end{equation} We have used $d$ rather than $D$ because we will see in a moment that we must compactify some of the flat dimensions of the previous subsection in order to get a consistent match of the entropy.  For large black holes, the Schwarzschild radius is at
\begin{equation} R^{d-1} = k_d E  l_P^{d - 2}R_{AdS}^2  .\end{equation}  Here $l_P$ is the $d$ dimensional Planck scale.   Using the area law for entropy, the entropy/energy relation is
\begin{equation} S(E) = k_d^{\prime} (E R_{AdS})^{\frac{d - 2}{d - 1}} (\frac{R_{AdS}}{l_P})^{\frac{d - 2}{d - 1}} . \end{equation}  This is the equation of state for a CFT on a $d - 2$ sphere of radius $\sim R_{AdS}$ but the coefficient in front of the energy dependence is larger than it would be for a single free field by a factor of order $(\frac{R_{AdS}}{l_P})^{\frac{d - 2}{d - 1}} $.

Our renormalization group transformation (the time evolution in the FRW patch at times of order $R_{AdS}$, acts on a space of cutoff fermionic degrees of freedom, whose Hilbert space at each fuzzy lattice point is of small dimension.  The UV limit of such a system will be a CFT of entropy of the same order as that of a small number of free fields.   Thus, there is a contradiction, in our HST construction of AdS space, between the explicit large, almost flat regime for proper times $t < R_{AdS}$, and our estimate of the entropy of the limiting CFT.  There is one, and we suspect only one, way to resolve this conundrum.  If $D > d$ and we take 
$D - d$ of the dimensions to be a compact manifold of volume $R_{AdS}^{D - d}$, then we get parametric agreement between the two estimates for entropy if
\begin{equation} (\frac{R_{AdS}}{l_P})^{\frac{d - 2}{d - 1}}  \lesssim R_{AdS}^{D - d}    . \end{equation} This is an inequality because some degrees of freedom in the cutoff system might be frozen out in the continuum limit.  The well known example of the RG flow from maximally supersymmetric $2 + 1$ dimensional Yang-Mills theory and the superconformal theory at the origin of its moduli space shows that in large $N$ theories we can lose entropy that is power law in $N$ upon taking the CFT limit.

To assess the meaning of this inequality in dimensions, we have to write $l_P$ in terms of $R_{AdS}$, recalling our convention that $L_P = 1$.  For $D - d$ compactified dimensions, the relation is
\begin{equation} l_P^{d - 2} \sim R_{AdS}^{d - D} . \end{equation}  Our previous inequality becomes
\begin{equation} R_{AdS}^{\frac{(d - D + 1)(d - 2)}{d - 1}}  \lesssim R_{AdS}^{D - d}    . \end{equation}  Since $R_{AdS}$ is large and positive this implies an inequality
\begin{equation} (D - d)(d - 1) \geq {(d - D + 1)(d - 2)} , \end{equation}
or
\begin{equation} D \geq d + \frac{d - 2}{2d - 3} . \end{equation}   We only have a sensible theory of $AdS_d$ for $d \geq 3$, and the idea of compactification implies $D \geq d + 1$ .  Thus, if we have AdS scale compactified dimensions, we will always satisfy this inequality, but $ d = D$ is inconsistent.   Note that it is also inconsistent to assume that the CFT we obtain in the limit of our HST model is a theory of free fermions, because we find that the gravitational calculation of the entropy is always less than the number of fermions, by a power of the AdS radius in D dimensional Planck units.  

The empirical evidence on large radius AdS/CFT compactifications suggests that $D - d \geq 2$.  We believe that the reason for this constraint has to do with {\it fast scrambling in the compact dimensions for large AdS black holes}.  This is to be rigorously differentiated from the conventional\cite{hpss} discussion of fast scrambling  on $S^{d - 2}$ in $AdS_d$, which does not occur for large black holes. Let us recall {\it why} large AdS black holes do not exhibit fast scrambling in the conventional sense.  Perturbations of the black hole can be divided according to whether the imaginary part of their quasi-normal mode frequency is larger or smaller than the natural temperature scale $T \gtrsim \frac{1}{R_{AdS}}$\cite{hhlfetal} .  Excitations with large imaginary frequencies dissipate exponentially rapidly, while those with small imaginary parts are hydrodynamic modes/quasi-particles of the dual field theory.  This is interpreted as rapid local thermalization of microscopic, non-hydrodynamic properties of the perturbation, followed by diffusive spread of hydrodynamic quantities over $S^{d-2}$.  The description of the system in terms of general relativity, does not allow us to {\it prove} that non-hydrodynamic information cannot be scrambled in a time $T^{-1} {\rm ln Entropy}$, but the dual field theory description proves this: the spread of information in any kind of lattice system is controlled by the Lieb-Robinson bound\cite{lr}.  Information can scramble at most ballistically $t_{scramble} \sim R$, with a system dependent velocity.  

On the other hand, the field theory description cannot prove a result for information spreading on the compact dimensions.  In all known models, the symmetries of the compact dimensions are internal symmetries, which act on the field space, rather than the space-time of the field theory.  Interactions are not local in field space, and in the cases where we can write a Lagrangian, they look like traces of products of large $N$ matrices that are in fixed representations of the isometry group of the internal manifold.  The bulk gravitational dual does provide evidence that excitations localized on the compact dimensions spread over those dimensions at the fast scrambling rate.  Indeed, Festuccia and Liu\cite{fl} have shown that perturbations in massive scalar fields have quasi-normal frequencies with an imaginary part that scales like $\left|{m}\right| T R_{AdS}$, if
$ \left|{m}\right|  R_{AdS} \gg 1$.  Eigenfunctions of the Laplacian on compact extra dimensions with radius of AdS scale give rise to AdS masses $m_L R_{AdS} \sim L$, where $L$ is the dimensionless eigenvalue.   A perturbation localized
in the compact dimensions on a scale $\delta \ll R_{AdS}$, will contain weight of order $1$ for eigenmodes with $m_L \sim \frac{1}{\delta} $, for which the calculation of \cite{fl} is valid.  This means that, up to exponentially small corrections, on the thermal time scale $T^{-1}$ the perturbation has only Laplacian eigenmodes whose maximal radius of localization is $o(R_{AdS})$.  This is fast scrambling on the internal space.  Note that this argument does not depend on the dimension or shape of the compact dimensions.

In HST, we can explain fast scrambling in $D - d \geq 2$ compact dimensions rather simply.  For $t < R_{AdS}$ the Hamiltonian $H_{in} (t)$ acts to rapidly scramble {\it all} the degrees of freedom, because it is invariant under fuzzy volume preserving mappings.  For larger times we must break this symmetry in the $d$ AdS directions in order to recover a local field theory as $t \rightarrow R_{AdS}$.  However, the labels on the fields in this cutoff field theory 
run over the fuzzy spinor bundle over the compact space and if we have two or more compact dimensions, then we can keep the fast scrambling form of the interactions in that index space.  In one dimension, this is not possible.  Note by the way that the latter observation is consistent with the fact that there are no black holes in $3$ dimensional Minkowski space.  

\section{Conclusions}

We have outlined a procedure interpolating between the AdS/CFT and HST descriptions of an AdS space with $R_{AdS} \gg L_S \sim 1$.  One notable feature of our analysis is an explanation of the necessity of large compact dimensions, whose volume is of order $R_{AdS_d}^{D - d}$ with $D - d \geq 2$.
The fact that the lower bound is $2$ follows from a new argument about the scrambling rate of information localized on the {\it compact} dimensions, for large AdS black holes.  We argued that the scrambling is characteristic of volume preserving mapping invariant, rather than local, Hamiltonians for dynamics on that geometry.  This follows from known properties of quasi-normal modes, and can be modeled in HST only if $D - d \geq 2$. 

A disturbing feature of our analysis is the relative disconnect between local physics on scales $\ll R_{AdS}$ and those features of the space-time which are well described by CFT.  It has of course been notoriously difficult to find even local operators smeared over the AdS scale in CFT, but if our analysis is correct it may be essentially impossible to extract much of sub-AdS physics from the CFT.  Generic low energy states of the CFT have a density matrix for the tensor factor of the Hilbert space that describes physics localized in a diamond with radius $\ll R_{Ads}$ which is extremely close to maximally uncertain.  Our description of local Minkowski scattering implies that such states have no localized excitations in any sub-AdS scale region.  Operators that construct the constrained states corresponding to localized excitations when acting on the AdS vacuum, have to destroy the maximal entanglement between the FRW patch Hilbert space and the rest of the CFT.

The authors of\cite{lennyjoeetal} have proposed a construction of such operators for small numbers of incoming and outgoing particles with energies well below the Planck scale, and the behavior of Feynman-Witten diagrams as the AdS radius goes to infinity certainly seems to support that conclusion.  It is however much less clear whether any AdS boundary measurement of less than horrendous complexity, could detect the fast scrambling behavior of small black holes, much less probe the question of whether black hole creation and decay in Minkowski space is described by a unitary S-matrix.

 \vskip.3in
\begin{center}
{\bf Acknowledgments }\\
The work of T.Banks is {\bf\it NOT} supported by the Department of Energy. The work of W.Fischler is supported by the National Science Foundation under Grant Number PHY-1316033. T.B. thanks S.Shenker for a very illuminating conversation about scrambling rates in quantum field theory.
\end{center}

\end{document}